\documentclass[]{pasj02} 
\usepackage[switch,mathlines]{lineno} 
\usepackage{natbib} 
\usepackage{url}
\usepackage{multirow}

\jyear{2024}
\Received{}
\Accepted{}

\begin{document} 

\title{Crimson Behemoth: a Massive Clumpy Structure Hosting a Dusty AGN at $z=4.91$}

\author{
Takumi S. \textsc{Tanaka},\altaffilmark{1,2,3} \email{takumi.tanaka@ipmu.jp} \orcid{0009-0003-4742-7060}
John D. \textsc{Silverman},\altaffilmark{1,2,3,4} \orcid{0000-0002-0000-6977}
Yurina \textsc{Nakazato},\altaffilmark{5} \orcid{0000-0002-0984-7713}
Masafusa \textsc{Onoue},\altaffilmark{2,6} \orcid{0000-0003-2984-6803}
Kazuhiro \textsc{Shimasaku},\altaffilmark{1,7} \orcid{0000-0002-2597-2231}
Yoshinobu \textsc{Fudamoto},\altaffilmark{8} \orcid{0000-0001-7440-8832}
Seiji \textsc{Fujimoto},\altaffilmark{9,10} \orcid{0000-0001-7201-5066}
Xuheng \textsc{Ding},\altaffilmark{11} \orcid{0000-0001-8917-2148}
Andreas L. \textsc{Faisst},\altaffilmark{12} \orcid{0000-0002-9382-9832}
Francesco \textsc{Valentino},\altaffilmark{10,13} \orcid{0000-0001-6477-4011}
Shuowen \textsc{Jin},\altaffilmark{10,14} \orcid{0000-0002-8412-7951}
Christopher C. \textsc{Hayward},\altaffilmark{15} \orcid{0000-0003-4073-3236}
Vasily \textsc{Kokorev},\altaffilmark{9,16} \orcid{0000-0002-5588-9156}
Daniel \textsc{Ceverino},\altaffilmark{17,18} \orcid{0000-0002-8680-248X}
Boris S. \textsc{Kalita},\altaffilmark{2,3,6} \orcid{0000-0001-9215-7053}
Caitlin M. \textsc{Casey},\altaffilmark{9,10} \orcid{0000-0002-0930-6466}
Zhaoxuan \textsc{Liu},\altaffilmark{1,2,3} \orcid{0000-0002-9252-114X}
Aidan \textsc{Kaminsky},\altaffilmark{19} \orcid{0009-0000-3672-0198}
Qinyue \textsc{Fei},\altaffilmark{20} \orcid{0000-0001-7232-5355}
Irham T. \textsc{Andika},\altaffilmark{21,22} \orcid{0000-0001-6102-9526}
Erini \textsc{Lambrides},\altaffilmark{23} \orcid{0000-0003-3216-7190}
Hollis B. \textsc{Akins},\altaffilmark{9} \orcid{0000-0003-3596-8794}
Jeyhan S. \textsc{Kartaltepe},\altaffilmark{24} \orcid{0000-0001-9187-3605}
Anton M. \textsc{Koekemoer},\altaffilmark{25} \orcid{0000-0002-6610-2048}
Henry Joy \textsc{McCracken},\altaffilmark{26} \orcid{0000-0002-9489-7765}
Jason \textsc{Rhodes},\altaffilmark{27} \orcid{0000-0002-4485-8549}
Brant E. \textsc{Robertson},\altaffilmark{28} \orcid{0000-0002-4271-0364}
Maximilien \textsc{Franco},\altaffilmark{9} \orcid{0000-0002-3560-8599}
Daizhong \textsc{Liu},\altaffilmark{29} \orcid{0000-0001-9773-7479}
Nima \textsc{Chartab},\altaffilmark{30} \orcid{0000-0003-3691-937X}
Steven \textsc{Gillman},\altaffilmark{10,14} \orcid{0000-0001-9885-4589}
Ghassem \textsc{Gozaliasl},\altaffilmark{31,32} \orcid{0000-0002-0236-919X}
Michaela \textsc{Hirschmann},\altaffilmark{33,34} \orcid{0000-0002-3301-3321}
Marc \textsc{Huertas-Company},\altaffilmark{35,36,37,38,39} \orcid{0000-0002-1416-8483}
Richard \textsc{Massey},\altaffilmark{40} \orcid{0000-0002-6085-3780}
Namrata \textsc{Roy},\altaffilmark{4} \orcid{0000-0002-4430-8846}
Zahra \textsc{Sattari},\altaffilmark{30,41} \orcid{0000-0002-0364-1159}
Marko \textsc{Shuntov},\altaffilmark{26} \orcid{0000-0002-7087-0701}
Joseph \textsc{Sterling},\altaffilmark{19} \orcid{0000-0002-2064-6429}
Sune \textsc{Toft},\altaffilmark{10,42} \orcid{0000-0003-3631-7176}
Benny \textsc{Trakhtenbrot},\altaffilmark{43} \orcid{0000-0002-3683-7297}
Naoki \textsc{Yoshida},\altaffilmark{2,5,7} \orcid{0000-0001-7925-238X}
Jorge A. \textsc{Zavala},\altaffilmark{44} \orcid{0000-0002-7051-1100}
}
\altaffiltext{1}{Department of Astronomy, Graduate School of Science, The University of Tokyo, 7-3-1 Hongo, Bunkyo-ku, Tokyo 113-0033, Japan}
\altaffiltext{2}{Kavli Institute for the Physics and Mathematics of the Universe (WPI), The University of Tokyo Institutes for Advanced Study, The University of Tokyo, Kashiwa, Chiba 277-8583, Japan}
\altaffiltext{3}{Center for Data-Driven Discovery, Kavli IPMU (WPI), UTIAS, The University of Tokyo, Kashiwa, Chiba 277-8583, Japan}
\altaffiltext{4}{Center for Astrophysical Sciences, Department of Physics and Astronomy, Johns Hopkins University, Baltimore, MD 21218, USA}
\altaffiltext{5}{Department of Physics, The University of Tokyo, 7-3-1 Hongo, Bunkyo, Tokyo 113-0033, Japan}
\altaffiltext{6}{Kavli Institute for Astronomy and Astrophysics, Peking University, Beijing 100871, China}
\altaffiltext{7}{Research Center for the Early Universe, Graduate School of Science, The University of Tokyo, 7-3-1 Hongo, Bunkyo-ku, Tokyo 113-0033, Japan}
\altaffiltext{8}{Center for Frontier Science, Chiba University, 1-33 Yayoi-cho, Inage-ku, Chiba 263-8522, Japan}
\altaffiltext{9}{Department of Astronomy, The University of Texas at Austin, 2515 Speedway Boulevard Stop C1400, Austin, TX 78712, USA}
\altaffiltext{10}{Cosmic Dawn Center (DAWN), Denmark}
\altaffiltext{11}{School of Physics and Technology, Wuhan University, Wuhan 430072, China}
\altaffiltext{12}{Caltech/IPAC, 1200 E. California Blvd. Pasadena, CA 91125, USA}
\altaffiltext{13}{European Southern Observatory, Karl-Schwarzschild-Str. 2, 85748, Garching, Germany}
\altaffiltext{14}{DTU Space, Technical University of Denmark, Elektrovej, Building 328, 2800, Kgs. Lyngby, Denmark}
\altaffiltext{15}{Center for Computational Astrophysics, Flatiron Institute, 162 Fifth Avenue, New York, NY 10010, USA}
\altaffiltext{16}{Kapteyn Astronomical Institute, University of Groningen, 9700 AV Groningen, The Netherlands}
\altaffiltext{17}{Departamento de Fisica Teorica, Modulo 8, Facultad de Ciencias, Universidad Autonoma de Madrid, 28049 Madrid, Spain}
\altaffiltext{18}{CIAFF, Facultad de Ciencias, Universidad Autonoma de Madrid, 28049 Madrid, Spain}
\altaffiltext{19}{Department of Physics, University of Miami, Coral Gables, FL 33124, USA}
\altaffiltext{20}{Department of Astronomy, School of Physics, Peking University, Beijing 100871, China}
\altaffiltext{21}{Technical University of Munich, TUM School of Natural Sciences, Department of Physics, James-Franck-Str. 1, 85748 Garching, Germany}
\altaffiltext{22}{Max-Planck-Institut für Astrophysik, Karl-Schwarzschild-Str. 1, 85748 Garching, Germany}
\altaffiltext{23}{NASA-Goddard Space Flight Center, Code 662, Greenbelt, MD, 20771, USA}
\altaffiltext{24}{Laboratory for Multiwavelength Astrophysics, School of Physics and Astronomy, Rochester Institute of Technology, 84 Lomb Memorial Drive, Rochester, NY 14623, USA}
\altaffiltext{25}{Space Telescope Science Institute, 3700 San Martin Drive, Baltimore, MD 21218, USA}
\altaffiltext{26}{Institut d’Astrophysique de Paris, UMR 7095, CNRS, and Sorbonne Université, 98 bis boulevard Arago, 75014 Paris, France}
\altaffiltext{27}{Jet Propulsion Laboratory, California Institute of Technology, 4800 Oak Grove Drive, Pasadena, CA 91001, USA}
\altaffiltext{28}{Department of Astronomy and Astrophysics, University of California, Santa Cruz, 1156 High Street, Santa Cruz, CA 95064, USA}
\altaffiltext{29}{Purple Mountain Observatory, Chinese Academy of Sciences, 10 Yuanhua Road, Nanjing 210023, China}
\altaffiltext{30}{The Observatories of the Carnegie Institution for Science, 813 Santa Barbara St., Pasadena, CA 91101, USA}
\altaffiltext{31}{Department of Computer Science, Aalto University, PO Box 15400, Espoo, FI-00 076, Finland}
\altaffiltext{32}{Department of Physics, Faculty of Science, University of Helsinki, 00014-Helsinki, Finland}
\altaffiltext{33}{Institute for Physics, Laboratory for Galaxy Evolution and Spectral Modelling, Ecole Polytechnique Federale de Lausanne, Observatoire de Sauverny, Chemin Pegasi 51, CH-1290 Versoix, Switzerland}
\altaffiltext{34}{INAF, Osservatorio Astronomico di Trieste, Via Tiepolo 11, I-34131 Trieste, Italy}
\altaffiltext{35}{Instituto de Astrofísica de Canarias (IAC), La Laguna 38205, Spain}
\altaffiltext{36}{Observatoire de Paris, LERMA, PSL University, 61 avenue de l’Observatoire, 75014 Paris, France}
\altaffiltext{37}{Université Paris-Cité, 5 rue Thomas Mann, 75014 Paris, France}
\altaffiltext{38}{Universidad de La Laguna, Avda. Astrofísico Fco. Sanchez, La Laguna, Tenerife, Spain}
\altaffiltext{39}{Center for Computational Astrophysics, Flatiron Institute, New York, USA}
\altaffiltext{40}{Institute for Computational Cosmology, Department of Physics, Durham University, South Road, Durham DH1 3LE, UK}
\altaffiltext{41}{Department of Physics and Astronomy, University of California, Riverside, 900 University Ave., Riverside, CA 92521, USA}
\altaffiltext{42}{Niels Bohr Institute, University of Copenhagen, Jagtvej 128, DK-2200 Copenhagen N, Denmark}
\altaffiltext{43}{School of Physics and Astronomy, Tel Aviv University, Tel Aviv 69978, Israel}
\altaffiltext{44}{National Astronomical Observatory of Japan, 2-21-1, Osawa, Mitaka, Tokyo, Japan}




\maketitle

\begin{abstract}
The current paradigm for the co-evolution of galaxies and their supermassive black holes postulates that dust-obscured active galactic nuclei (AGNs) represent a transitional phase towards a more luminous and unobscured state.
However, our understanding of dusty AGNs and their host galaxies at early cosmic times is inadequate due to observational limitations.
Here, we present JWST observations of CID-931, an X-ray-detected AGN at a spectroscopic redshift of $z_{\rm spec}=4.91$.
Multiband NIRCam imaging from the COSMOS-Web program reveals an unresolved red core, similar to JWST-discovered dusty AGNs.
Strikingly, the red core is surrounded by at least eight massive star-forming clumps spread over $1.\!\!^{\prime\prime}6 \approx 10~{\rm kpc}$, each of which has a stellar mass of $10^9-10^{10}M_\odot$ and $\sim0.1-1~{\rm kpc}$ in radius.
The whole system amounts to $10^{11}M_\odot$ in stellar mass, higher than typical star-forming galaxies at the same epoch.
In this system, gas inflows and/or complex merger events may trigger clump formation and AGN activity thus leading to the rapid formation of a massive galaxy hosting a supermassive black hole.
Future follow-up observations will provide new insights into the evolution of the galaxy-black hole relationship during such transitional phases in the early universe. 
\end{abstract}


\section{Introduction}\label{s:intro}

The James Webb Space Telescope (JWST, \citealt{Rigby2023}) has opened explorations of the distant universe.
JWST has discovered many massive galaxies with stellar masses of $M_* \sim 10^{11} M_\odot$ at $z \gtrsim 4$ (e.g., \citealt{Chworowsky2023, deGraaff2024}) and $M_* \sim 10^{10} M_\odot$ even at $z \gtrsim 10$ \citep{Casey2024}.
Observational studies have also identified hundreds of quasars with high black hole masses ($M_{\rm BH} \gtrsim 10^8 M_\odot$) at $z \gtrsim 5$ (e.g., \citealt{Fan2023}).
However, the rapid growth mechanism(s) of galaxies and black holes in the early universe remains a key unresolved problem. The distant universe is likely to provide important insights into the observed relations between the properties of supermassive black holes (SMBHs) and galaxies such as stellar mass $M_*$, bulge mass $M_{\rm bulge}$, and stellar velocity dispersion $\sigma_*$ as seen at low-$z$ (e.g., \citealt{Magorrian1998, Kormendy2013, Reines2015}). 

In the galaxy and SMBH co-evolution paradigm (e.g., \citealt{Hopkins2008}), dusty active galactic nuclei (AGNs) are considered a transitional phase with AGN feedback expelling the dust to reveal an unobscured luminous quasar phase (e.g., \citealt{Gilli2011, Vito2019, Kato2020, Fujimoto2022}). 
As well, such feedback is considered to impact the interstellar medium (ISM), suppress star formation, and eventually lead to a massive quiescent galaxy.

Despite the importance of obscured AGNs in the lower redshift universe, relatively less is known about high-$z$ obscured AGNs and their host galaxies.
This gap results in missing pieces in our understanding of galaxy evolution in the early universe including the triggers of AGN activity and the processes driving the evolution of their host galaxies.

Recently, JWST has identified a new high-$z$ galaxy population named ``little red dots'' (LRDs), characterized by their compactness and red color in rest-frame optical wavelengths (e.g., \citealt{Labbe2023, Furtak2023, Kocevski2023, Harikane2023, Akins2023, Barro2024, Matthee2024, Labbe2023a, Akins2024}). 
In many cases, by definition, no extended components are detected even after point spread function (PSF) subtraction thus truly having a compact morphology which JWST cannot resolve ($\lesssim 200~{\rm pc}$).

The nature of LRDs is still debated: whether LRDs are dusty star-forming galaxies or dusty AGNs (e.g., \citealt{Barro2024, Labbe2023, Perez2024, Ananna2024, Akins2024, Kokorev2024_bb}).
JWST spectroscopic observations have confirmed broad emission lines (full width at half maximum (FWHM) $> 1000~{\rm km~{s^{-1}}}$) in photometrically selected LRDs with high completeness (e.g., \citealt{Greene2024} reported that 80\% of objects with $m_{\rm F277W}-m_{\rm F444W}>1.5$ have broad line feature), suggesting that LRDs are dusty type-I AGNs (e.g., \citealt{Kocevski2023, Harikane2023, Matthee2024, Furtak2024, Kokorev2023, Greene2024, Kocevski2024}).
Another notable feature is that the number density of LRDs (e.g., \citealt{Kokorev2023, Kocevski2024, Akins2024}) and high-$z$ AGNs (e.g., \citealt{Onoue2023, Harikane2023, Maiolino2023, Scholtz2023}) is higher than an extrapolation from the luminosity function based on the high-$z$ quasar studies prior to JWST (e.g., \citealt{Matsuoka2018, Niida2020}).

Of particular importance, constraining the stellar mass ($M_*$) of LRDs is challenging due to their compactness and difficulties in decomposing their spectral energy densities (SEDs). Upper limits on $M_*$ suggest $M_{\rm BH}/M_*$ ratios higher than the local mass relation thus revealing overmassive black holes (e.g., \citealt{Harikane2023, Kokorev2023, Killi2023, Furtak2024}).
The interpretation of such overmassive high-$z$ SMBHs in the co-evolution scheme — whether they represent the intrinsic SMBH population (e.g., \citealt{Pacucci2023, Durodola2024}) or a subset due to selection biases (e.g., \citealt{Ding2022_z6, Li2024})—continues to be debated.
The compactness selection criteria used in previous LRD studies may also bias the selected galaxies toward AGN-dominant systems lacking extended host galaxy components.

\begin{figure*}
 \begin{center}
  \includegraphics[width=17cm]{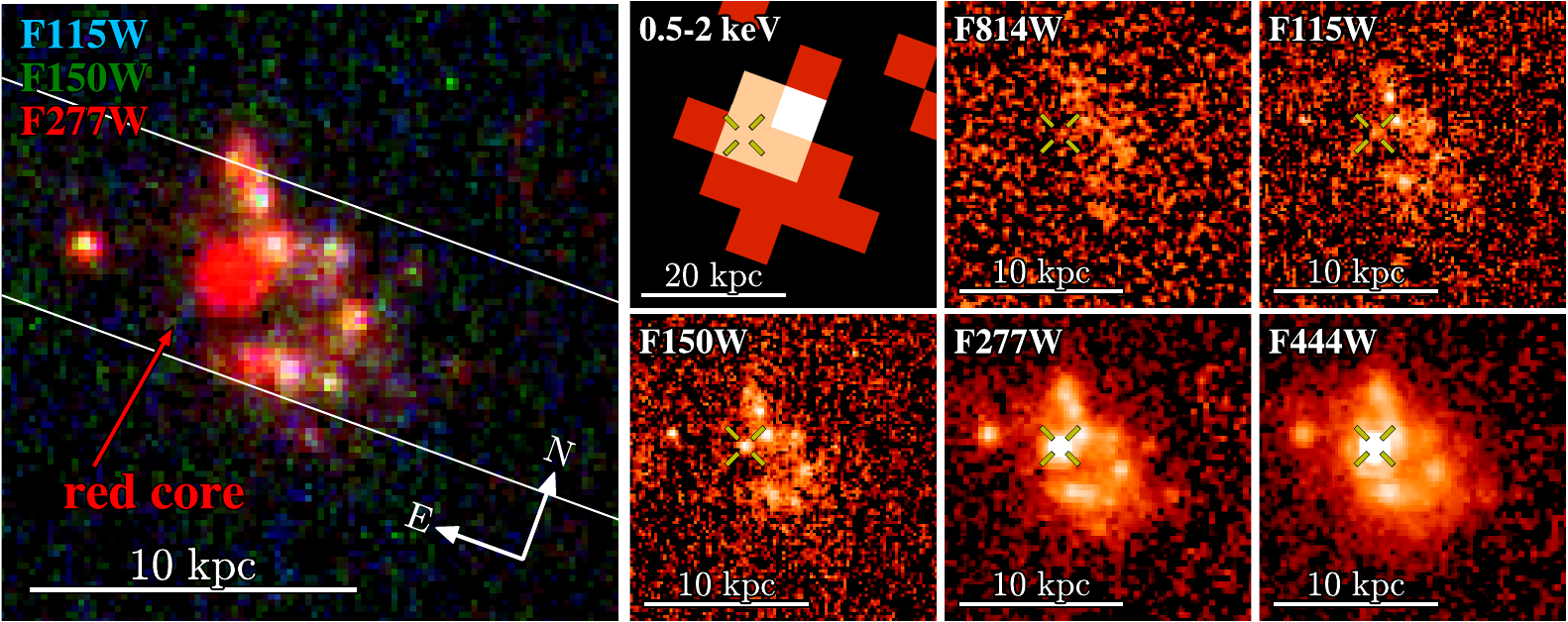} 
 \end{center}
 \caption{Three-color (JWST/NIRCam F277W, F150W, and F115W for RGB) image and separate filter images of CID-931.
 The slit position of Keck-II/DEIMOS observation is shown by the white lines. 
 Note that the Chandra 0.5-2~keV image has a different cutout size ($6^{\prime\prime}\times6^{\prime\prime}$) due to a much lower pixel scale ($0\farcs984/{\rm pixel}$) than the HST and JWST images ($3^{\prime\prime}\times3^{\prime\prime}$ cutout with the pixel scale of $0\farcs03/{\rm pixel}$).
 Yellow bars in right panels indicate the position of the red core.
 }
\label{fig;imgs}
\end{figure*}

To explore high-$z$ dusty AGNs free from the aforementioned possible biases, we search for LRDs-like colored objects with extended host galaxy components using COSMOS-Web, a JWST Cycle 1 large program \citep{Casey2023}.
The morphology and structure of the underlying host galaxies may shed light on their formation, such as signs of mergers, interactions, or more secular processes.
This paper reports the image-based analysis of an exceptional case, CID-931\footnote{Dubbed ``Crimson Behemoth'' due to showing a red core, possibly a dusty AGN, and a massive galaxy with an unprecedented clumpy structure.}, an X-ray-detected AGN at $z_{\rm spec}\sim4.91$ with an unresolved red core and an extended highly clumpy structure at rest-frame optical wavelengths.
The remainder of this paper is organized as follows.
The data and the target are described in section~\ref{s:data}.
Section~\ref{s:me_re} introduces the image-based decomposition and SED fitting analysis.
Then, we present the AGN- and galaxy-related results in section~\ref{s:AGN}, respectively.
Finally, we discuss how the nature of CID-931 relates to LRDs in the literature (Section 5) and conclude with our results in section~\ref{s:conclusion}.
Throughout this paper, the AB magnitude system \citep{Oke1983} is adopted, and we assume a \citet{Chabrier2003} initial mass function and a standard cosmology with $H_0 = 70~{\rm km~s^{-1}~Mpc^{-1}}$, $\Omega_m = 0.30$, and $\Omega_\Lambda=0.70$.

\section{Data and target selection}\label{s:data}

\subsection{Data}
COSMOS-Web (PI: Jeyhan Kartaltepe and Caitlin Casey, GO1727, \citealt{Casey2022} for the overview) is a JWST treasury survey program observed in Cycle~1 with wide-area coverage of 0.54~${\rm deg}^2$ using NIRCam (\citealt{Rieke2023}; F115W, F150W, F277W, F444W) and 0.19 ${\rm deg}^2$ with MIRI (\citealt{Bouchet2015}; F770W).
CID-931 is only covered by the four NIRCam filter images from COSMOS-Web.
The data is reduced with the JWST Calibration Pipeline\footnote{\url{https://github.com/spacetelescope/jwst}} \citep{jwst_pipeline} version 1.10.0 and the calibration Reference Data System version 1075.
The final pixel scale is 0\farcs030/pixel (see Franco et al., in preparation for details of the image reduction).

In addition to the NIRCam images, we use data from Chandra/ACIS-I (COSMOS Chandra Merged Image Data Version 1.0, \citealt{Civano2016}\footnote{\url{https://irsa.ipac.caltech.edu/data/COSMOS/images/chandra/merged/}}), Hubble Space Telescope HST/ACS (\citealt{Koekemoer2007}; F814W), Subaru Hyper Suprime-Cam (HSC Subaru Strategic Program DR3, \citealt{Aihara2022}; $grizy$), and Spitzer/IRAC and MIPS (S-COSMOS, \citealt{Sanders2007}; 3.6, 4.5, 5.6, 8.0, 24.0~$\mu$m).
In this study, we also use the photometric information from the XMM-Newton COSMOS X-Ray Point Source Catalog \citep{Cappelluti2009} and the Super-deblended catalog (\citealt{Jin2018}; Hershel/PACS $100{\rm \mu m}$, $160{\rm \mu m}$, Hershel/SPIRE $250{\rm \mu m}$, $350{\rm \mu m}$, $500{\rm \mu m}$, $850{\rm \mu m}$, and JCMT/SCUBA2 $850~{\rm \mu m}$).

\subsection{Identification of CID-931, an X-ray-detected AGN at $z=4.91$}\label{ss:sample}

\begin{figure}
 \begin{center}
  \includegraphics[width=8cm]{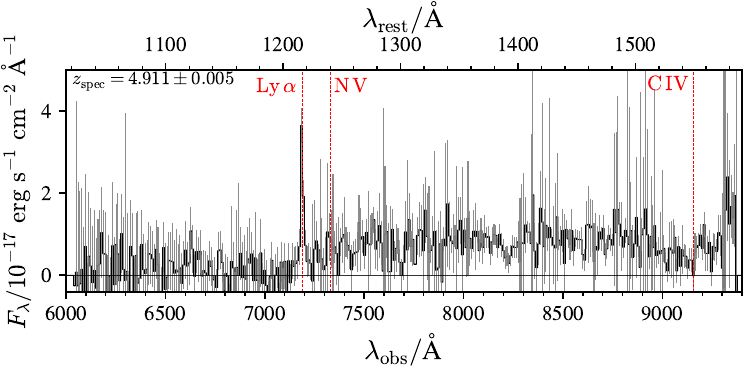} 
 \end{center}
 \caption{Keck II/DEIMOS spectrum \citep{Hasinger2018} binned over 20 spectral elements to increase the $S/N$.
 Gray lines indicate $1\sigma$ error of each spectral bin.
 The positions of Ly$\alpha$, N\,{\sc v} $\lambda1240$, and C\,{\sc iv} $\lambda 1549$ are indicated by vertical red lines.
 }
\label{fig;spec}
\end{figure}

Previous LRD studies use color and morphological information to select high-$z$ AGN candidates from JWST imaging data (e.g., \citealt{Labbe2023a, Labbe2023, Barro2024, Greene2024, Kocevski2024, Akins2024}).
\cite{Barro2024} employed a single color selection of $m_{\rm F277W} - m_{\rm F444W} > 1.5$ and found 37 LRDs.
\cite{Greene2024} spectroscopically confirmed broad Balmer emission lines from such photometrically selected LRDs and suggested that a color selection of $m_{\rm F277W} - m_{\rm F444W} > 1.6$ can effectively select broad-line LRDs with a low contamination fraction.
\cite{Kocevski2024} used a similar strategy, in which they performed continuum slope fitting to select LRDs over a wide redshift range. 

To avoid a bias towards AGN-dominated systems, we apply a single color selection of $m_{\rm F277W} - m_{\rm F444W} > 1.5$ to the flux in each pixel separately (a pixel-by-pixel color selection, see Tanaka et al., in preparation for details of the selection method). Our candidate list is then cross-matched to the COSMOS DEIMOS spectroscopic catalog \citep{Hasinger2018}, which includes 10,718 objects using the Deep Imaging Multi-Object Spectrograph (DEIMOS, \citealt{Faber2003}) on the Keck II telescope. 
This results in the identification of CID-931 which was targeted due to being an X-ray source. The DEIMOS spectrum provided the redshift $z_{\rm spec} = 4.911\pm0.005$ based on the $6.7\sigma$ detection of Ly$\alpha$ line.

CID-931 was detected by Chandra and included in the COSMOS Legacy Survey Point Source Catalog \citep{Civano2016} based on a 4.6~Ms Chandra X-ray observations on the COSMOS field, which strongly suggests the presence of an AGN. We note that \cite{Li2024} independently reported CID-931 as an offset AGN, in which AGNs and their host galaxies merge with non-active galaxies.

As shown in the top left panel of figure~\ref{fig;imgs}, CID-931 has a bright red core surrounded by a highly clumpy structure of various colors and extending up to $\sim10$~kpc (1\farcs6) in diameter. The DEIMOS slit covers almost all of the clumps (figure~\ref{fig;spec}). While the individual clumps cannot be spatially resolved in the Subaru HSC images, CID-931 is observed as a single extended component with Sérsic index of $n\sim1$ and effective radius of $r_e\sim0\farcs5$ (corresponding to $\sim3~{\rm kpc}$) at wavelengths longer than the $i$-band.
The whole system of CID-931 is undetected in HSC $g$- and $r$-band, consistent with its spectroscopic redshift of $z_{\rm spec}=4.91$. Thus, we conclude that the clumpy structure is not a foreground object such as a star cluster in our Galaxy.  These clumps show variety in their colors (see section~\ref{ss:clumps}); therefore, they are unlikely to be gravitationally-lensed multiple images.

CID-931 was not detected in \cite{Akins2024}, which utilized compactness and $m_{\rm F277W} - m_{\rm F444W} > 1.5$ color selection and identified 447 LRDs from COSMOS-Web data.
If we measure aperture photometry by centering on the red core, it can pass the compactness criteria with $F_{\rm F444W,~0\farcs2}/F_{\rm F444W,~0\farcs5} = 0.55$.
However, the red core does not pass the color selection with $m_{\rm F277W} - m_{\rm F444W} = 1.1$ due to the contamination from a nearby clump.
We note that the photometry of the red core alone, estimated from the image-based analysis (section~\ref{ss:decomposition}), passes the color selection criteria (section~\ref{ss:agn}).

\section{Analysis}\label{s:me_re}

\subsection{Image-based decomposition}\label{ss:decomposition}
Light from an unresolved AGN usually outshines the stellar emission from its host galaxy.
Therefore, image-based decomposition methods are needed to extract information about the host galaxies (e.g., \citealt{Ding2020_HST, Ding2022_z6, Zhuang2023COSMOS, Tanaka2024, Kocevski2024}).
In this approach, the imaging data is fit with a 2D model that combines a PSF component representing the AGN and a Sérsic profile(s) \citep{sersic1968}, convolved with the point spread function (PSF), representing the extended host galaxy.
We can then remove the AGN emission by subtracting the PSF component from the original image.

To decompose the highly clumpy structure of CID-931 and obtain the flux of the red core component, we fit the imaging data from HST and JWST with a model composed of multiple components using the image analysis tool {\tt galight} \citep{Ding2020_HST}.
First, we perform automatic segmentation on the F150W image, which offers a moderate signal-to-noise ratio $S/N$ and spatial resolution compared to the other bands.
We utilize the {\tt detect\_obj} function in {\tt galight} to select regions with $S/N >1.5$ that extend over more than 5 pixels.
Then, we manually remove false detections from the automated segmentation map that are undetected in the other bands. 
We finally separate the system into 13 components and the red core.

We use the position of each segmented component to construct a composite model.
All components except the red core are fitted with single Sérsic profiles.
For the red core, we attempt three different models: a PSF, a single Sérsic profile, and a PSF + a single Sérsic profile.
Note that the red core and the other clumps are fitted simultaneously.
We limit the effective radii $r_e$ and Sérsic indices $n$ to [0\farcs01, 0\farcs3] and [0.3, 9.0], respectively.
The center position of each component is allowed to vary between different images within a range of 0\farcs1 around the initial estimates based on the F150W segmentation map.

Figure~3 presents our best model that reproduces the observed NIRCam images well.
Comparing the fitting results using the three different models for the red core, the Bayesian Information Criterion (BIC) values indicate that ${\rm BIC_{PSF}}$ is less than ${\rm BIC_{S\acute{e}rsic}} - 10$ and ${\rm BIC_{PSF+S\acute{e}rsic}} - 10$ in all bands\footnote{For the thresholds for the BIC differences, we use 10 based on \cite{Kass1995}.}, suggesting that the PSF component adequately describes the red core profile.
Even using the Sérsic profile for the red core, a best fit $r_e$ is $\sim0\farcs04$ ($\sim0.3~{\rm kpc}$) in F150W, indicating a very compact morphology.
Therefore, we derive the flux of the red core and the total system from the fitting results using the PSF component for the red core.
The red core is not significantly detected in F814W ($S/N=1.8$), and we use its 3$\sigma$ upper limit ($=0.030~{\rm \mu Jy}$) in the following SED analysis.

\begin{figure}
 \begin{center}
  \includegraphics[width=8cm]{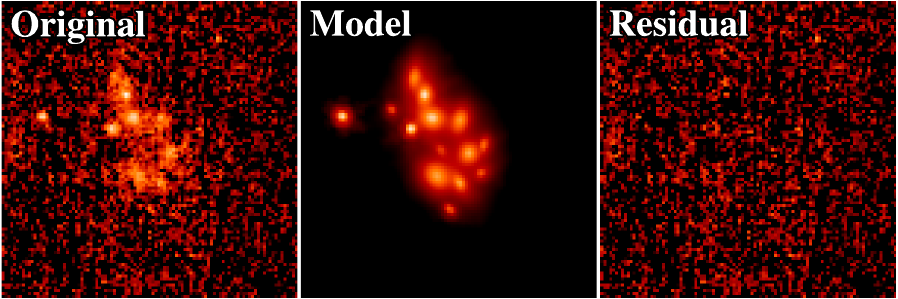} 
 \end{center}
\caption{
Image-based decomposition analysis in F150W filter:
(Left) original image of CID-931 ($3^{\prime\prime}\times3^{\prime\prime}$),
(Middle) best-fit model composed of a PSF component for the red core and Sérsic profiles for the other components, and 
(Right) residual image, i.e., left minus right.
}
\label{fig;decomposition}
\end{figure}

\subsection{SED fitting analysis}\label{ss:sed_fitting}
We perform an SED fitting analysis for the whole system of CID-931 based on its total photometry from rest-frame X-ray to far-infrared wavelengths.
For X-ray photometry, we utilize data from the Chandra Legacy Point Source Catalog \citep{Civano2016}.
For HSC and Spitzer, we estimate the photometry after subtracting the contribution from a nearby star. 
We use the total photometry obtained from 2D decomposition results (section~\ref{ss:decomposition}) for HST and JWST.
Because the Herschel and JCMT/SCUBA2 FIR photometry from the Super-deblended catalog \citep{Jin2018} have $S/N\lesssim2$, we use $3\sigma$ values as upper limits in the fitting.

We input these photometric data into CIGALE \citep{Boquien2019, Yang2022} and fit them using a galaxy plus {\tt skirtor2016} AGN template \citep{Stalevski2012, Stalevski2016}.
For the stellar population model, we use the \cite{Bruzual2003} model ({\tt bc03} module) and assume the initial mass function by \cite{Chabrier2003}.
For the star-formation history, we assume a delayed-$\tau$ model which equalizes a star formation rate (SFR) at each look-back time $t$ as,
\begin{equation}
    {\rm SFR}\left(t\right) \propto
    \begin{cases}
        \left(t-t_{\rm age}\right) \exp\left(-\frac{t-t_{\rm age}}{\tau}\right) & \left(t>t_{\rm age}\right),\\
        0 & \left(t<t_{\rm age}\right),
    \end{cases}
\end{equation}
where $t_{\rm age}$ and $\tau$ are the starting time of star-formation activity and the SFR declining timescale.
We also include nebular emission with {\tt nebular} module and fix the ionized parameter as $\log\,U=-2$.
For dust attenuation, we utilize a {\tt dustatt\_modified\_starburst} module that assumes the modified \cite{Calzetti2000} law.
The detailed parameter settings are summarized in table~\ref{tab:sed}.

\begin{table*}[]
\scriptsize
\caption{Major parameter settings in CIGALE SED fitting}\label{tab:sed}
\begin{tabular}{llp{4cm}p{4cm}}
\hline\hline
module & parameter & description & values\\
\hline
{\tt sfhdelayed} & {\tt tau\_main} & timescale of delayed-$\tau$ SFH in Myr & 50, 100, 250, 500, 750, 1000, 2000\\
& {\tt age\_main} & starting time of delayed-$\tau$ SFH in Myr & 50, 100, 250, 500, 750, 1000\\
\hline
{\tt bc03} & {\tt imf} & initial mass function & 1: \cite{Chabrier2003}\\
& {\tt metallicity} & stellar metallicity & 0.02 (solar metallicity)\\
\hline
{\tt nebular} & {\tt logU} & ionization parameter & -2\\
\hline
{\tt dustatt\_modified\_starburst} & {\tt E\_BV\_lines} & color excess for the nebular lines & 0, 0.05, 0.1, 0.3, 0.5\\
& {\tt powerlaw\_slope} & slope delta of the power law modifying the attenuation curve & 0.0, 0.5, 1.0 \\
\hline
{\tt skirtor2016} & {\tt t} & AGN average edge-on optical depth at 9.7~${\rm \mu m}$ & 3, 7, 11\\
& {\tt i} & viewing angle in deg & 10, 30, 70\\
& {\tt delta} & power-law index for modifying the disk SED & -0.3, 0, 0.3\\
& {\tt fracAGN} & AGN contribution to IR luminosity & 0.1, 0.2, 0.3, 0.4\\
& {\tt EBV} & polar-dust color excess & 0, 0.05, 0.1, 0.3, 0.5, 0.75, 1.0 \\
\hline
\end{tabular}
\end{table*}

We also implement a ``pixel-by-pixel'' SED fitting method, which performs SED fitting on photometry in each pixel using only HST and JWST imaging data.
In this analysis, we first cutout the $3^{\prime\prime}\times3^{\prime\prime}$ region to cover all clumps around the red core. 
We match the PSF sizes of the HST and JWST cutout images after subtracting the PSF component of the red core.
For this procedure, we use F444W as a reference image, as its PSF is the largest among those of the images used.
The images are binned by $2\times2$ pixel in order to increase the $S/N$ in each pixel, and the final pixel scale is $0\farcs060/{\rm pixel}$ that is still smaller than the PSF FWHM of F444W ($0\farcs14 - 0\farcs15$).
Then, we only use the binned pixels with $S/N > 1$ in F150W in the fitting.
Because we only use data of five rest-UV to rest-optical filters of HST and JWST, and the red core PSF component has been subtracted, the AGN component is not included in the SED fitting.

\section{Results}\label{s:AGN}

As detailed below, CID-931 is an X-ray source confirmed to be at a spectroscopic redshift of $z_{\rm spec}=4.91$ based on its rest-UV spectrum (section~\ref{ss:spec}).
The X-ray emission is luminous at this redshift that can only be explained by an AGN (section~\ref{ss:xray}).
The red core shares a similar optical slope with known distant dusty AGNs and is likely responsible for the X-ray emission (section~\ref{ss:agn}).
Additionally, unlike other known high-$z$ dusty AGNs, the red core is uniquely surrounded by multiple, massive star-forming clumps (section~\ref{ss:host}) which lends insight into its rapid assembly.

\subsection{The Rest-UV spectrum}\label{ss:spec}
The Keck II DEIMOS spectrum (figure~\ref{fig;spec}) exhibits a Ly$\alpha$ emission line.
From the single Gaussian + constant model fitting around the Ly$\alpha$, we confirm that the Ly$\alpha$ line of CID-931 is a narrow line with the FWHM of $\sim 500~{\rm km~s^{-1}}$.
The Ly$\alpha$ obscured (not attenuation-corrected) line luminosity is $\sim 1.2\times10^{44}~{\rm erg~s^{-1}}$ that is around the bright end of the high-$z$ Ly$\alpha$ luminosity function (e.g., \citealt{Konno2018, Goto2021, Morales2021}). 
Considering this line luminosity, the Ly$\alpha$ line contributes $\sim 0.4~{\rm \mu Jy}$ to the F814W flux density, and it significantly exceeds the $3\sigma$ upper limit for the F814W flux density of the red core ($0.03~{\rm \mu Jy}$).
Therefore, we consider that the Ly$\alpha$ line emission mainly comes from the star-forming clumps rather than the red core, which is the most likely AGN component of the CID-931 system.
C\,{\sc iv} $\lambda 1549$, a typically-observed high-ionized line in AGN spectra, is undetected.
Significant dust attenuation in the red core can bury Ly$\alpha$ and C\,{\sc iv} lines from the AGN.

\subsection{X-ray emission}\label{ss:xray}
CID-931 is detected by Chandra in the soft band (0.5--2 keV) with $S/N=4.4$ and full band (0.5--7 keV) with $S/N=3.9$.
However, CID-931 is not detected in the hard band (2--7 keV).
CID-931 is also detected by XMM-Newton in the soft band (0.5--2 keV) with $S/N\sim6$ and not detected in the hard band (2--10 keV). 
The soft-band flux from XMM-Newton $\left(1.2\pm0.2\right)\times10^{-15}~{\rm erg~s^{-1}~cm^{-2}}$ is consistent with the flux from Chandra $\left(0.9\pm0.2\right)\times10^{-15}~{\rm erg~s^{-1}~cm^{-2}}$ within the $1\sigma$ uncertainty.

Because the X-ray net counts ($25.5\pm6.9$ counts) are insufficient for spectral analysis, it is challenging to strongly constrain parameters such as hydrogen column density $N_{\rm H}$ and photon index $\Gamma$.
However, due to non-detection in the hard band, we assign an upper limit for the hard band flux (90\% confidence level estimated in \citealt{Civano2016}, $1.5\times10^{-15}~{\rm erg~cm^{-2}~s^{-1}}$) and obtain an intrinsic $N_{\rm H}$ upper limit as $N_{\rm H}\lesssim8\times10^{22}~{\rm cm^{-2}}$, assuming a photon index $\Gamma=1.9$ (e.g., \citealt{Marchesi2016, Ricci2017}).
When using the XMM-Newton soft-band flux and the $3\sigma$ upper limit for the hard-band flux ($5\times10^{-15}~{\rm erg~cm^{-2}~s^{-1}}$), we obtain an upper limit of $N_{\rm H}$ as $N_{\rm H}\lesssim7\times10^{23}~{\rm cm^{-2}}$.
These upper limits are far from the threshold for being a Compton-thick AGN ($N_{\rm H}\gtrsim10^{24}~{\rm cm^{-2}}$).
Therefore, we can rule out the possibility that CID-931 is a Compton-thick AGN.

Considering the intrinsic $N_{\rm H}$ range of $0 < N_{\rm H}/{\rm cm^{-2}} \lesssim 8\times10^{22}$ and accounting for the Galactic $N_{\rm H}$ (\citealt{HI4PI2016}, retrived from NASA HEASARC tools\footnote{\url{https://heasarc.gsfc.nasa.gov/cgi-bin/Tools/w3nh/w3nh.pl}}), we obtain the X-ray unobscured luminosity of $2.6\times10^{44} < L_{[2-10~{\rm keV}]}/{\rm erg~s^{-1}} < 3.2\times10^{44}$.
These X-ray unobscured luminosity cannot be attributed to only X-ray binaries; e.g., the relation between normal galaxy X-ray luminosity, $M_*$, and SFR \citep{Lehmer2010} requires $M_*\sim10^{16}M_\odot$ or ${\rm SFR}\sim10^5 M_\odot~{\rm yr}^{-1}$ to describe the observed X-ray unobscured luminosity. 
Therefore, the X-ray emission from CID-931 is securely due to an AGN.

\subsection{The red core as a dusty AGN}\label{ss:agn}

With the red core well fit by a PSF model and being the brightest component in CID-931, the simple interpretation is that the red core is the probable site of the X-ray source.
However, the Chandra X-ray image (figure~\ref{fig;imgs}) lacks sufficient spatial resolution (size of 50\% enriched energy fraction is about $2-4^{\prime\prime}$, \citealt{Civano2016}) to uniquely confirm which component in the CID-931 system corresponds to the X-ray source.

\begin{figure}
 \begin{center}
  \includegraphics[width=8cm]{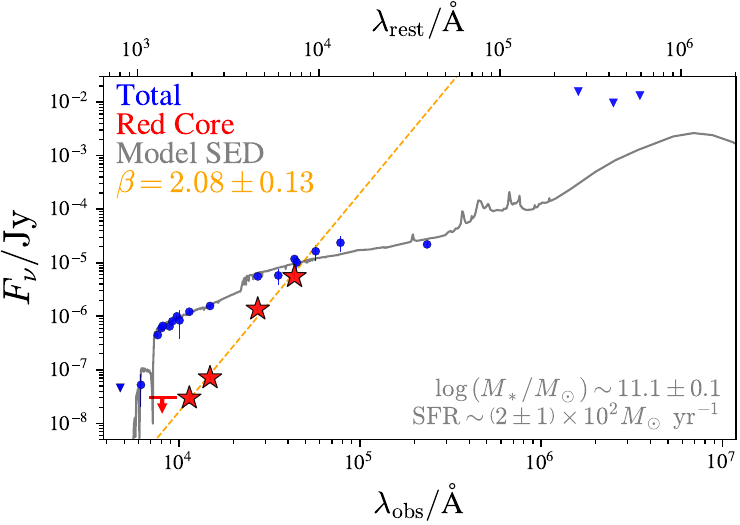} 
 \end{center}
\caption{
SED of the full CID-931 system (blue) and the red core separately (red).
Triangle markers indicate the $3\sigma$ upper limit.
The orange line indicates the best-fit power law of F115W to F444W photometry with $\beta=2.08\pm0.13$.
The gray line indicates the best-fit model SED.
}
\label{fig;core_SED}
\end{figure}

To assess the color of the red core, we fit the photometry from F115W to F444W by a single power-law SED model.
We find that the spectral slope is $\beta = 2.08 \pm 0.13$, where $\beta$ is defined as $F_\lambda \propto \lambda^\beta$.
The slope is significantly redder than that of a typical unobscured quasar ($\beta=-1.5$; e.g., \citealt{Vanden2001}), and that of the whole system of CID-931 (figure~\ref{fig;core_SED}). 
The color index of the red core, $m_{\rm F277W}-m_{\rm F444W}=1.5~{\rm mag}$ (2.1~mag when calculated from $\beta$) indicates that it is as red as high-$z$ dusty AGNs (e.g., \citealt{Lambrides2024} reported $m_{\rm F277W}-m_{\rm F444W}=1.8~{\rm mag}$) and spectroscopically confirmed broad-line LRDs (e.g., \citealt{Greene2024}). Additionally, the F277W photometry shows a $2.8\sigma$ excess ($-0.35~{\rm mag}$) over the fitted power law.
This F277W excess potentially indicates contributions from H$\beta$ + [O\,{\sc iii}] $\lambda\lambda{\rm 4959, 5007}$ ($\lambda_{\rm obs} = 2.87,~2.93,~2.96~{\rm \mu m}$, respectively), similar to other spectroscopically confirmed JWST AGN \citep{Onoue2023, Kocevski2023, Solimano2024}.

We also perform SED fitting for just the red core using the galaxy SED models without an AGN model.
In this case, we recover a stellar mass of $10^{11.7} M_{\odot}$, which is extremely higher than typical $z=5$ massive galaxies (e.g., \citealt{Weaver2023}).
Moreover, the compact size of the red component yields the stellar mass density of $\Sigma_* \gtrsim 3\times10^{12}M_\odot~{\rm kpc^{-2}}$.
This is also significantly higher than the maximum $\Sigma_*$ ever found ($\Sigma_*\sim 3\times10^{11}M_\odot~{\rm kpc}^{-2}$, \citealt{Hopkins2010, Grudic2019}).
Whereas fitting with an AGN plus a galaxy composite model results in $\log\left(M_*/M_\odot\right) = 10.9 \pm 0.2$, accounting for $\sim60\%$ of the total $M_*$ (section~\ref{ss:host}). This analysis is supportive of the red core including a dusty AGN.

\begin{figure}
 \begin{center}
  \includegraphics[width=8cm]{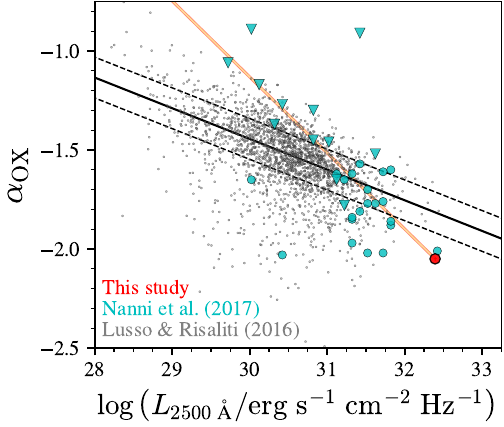} 
 \end{center}
\caption{
The relation between $L_{2500}$ and $\alpha_{\rm OX}$. 
Red, cyan, and gray circles indicate CID-931, high-$z$ ($z\sim6$) luminous quasars in \cite{Nanni2017}, and low-to-mid-$z$ (largely $z\lesssim3$) AGNs in \cite{Lusso2016}.
Black solid and dashed lines indicate the relation by \cite{Lusso2016} and its 1$\sigma$ interval.
Triangle markers indicate the upper limit of $\alpha_{\rm OX}$.
An orange-shaded line indicates the potential change in $\alpha_{\rm OX}$ depending on the decrease in $L_{2500}$ due to the uncertain host galaxy contribution and dust extinction.
}
\label{fig;alphaOX}
\end{figure}

Assuming the SED of the red core to be entirely due to an AGN, we fit the HST and JWST photometry to determine the continuum spectral slope ($f_\lambda \propto \lambda^\beta$, e.g., \citealt{Vanden2001}). We find $\beta=-1.5$ and then estimate the unobscured rest-2500~\AA~monochromatic luminosity $L_{2500}$ considering dust attenuation \citep{Calzetti2000}.
This results in $A_V=2.6\pm0.3~{\rm mag}$, and $L_{2500}=2.5^{+0.4}_{-0.3}\times10^{32}~{\rm erg~s^{-1}~{cm^{-2}~{Hz^{-1}}}}$.
As shown in figure~\ref{fig;alphaOX}, the value of $L_{2500}$ is similar to very UV-bright quasars \citep{Lusso2016, Nanni2017}.

With the X-ray luminosity, we obtain $\alpha_{\rm OX}\sim-2.0$, which is the ratio of the X-ray (2~keV) monochromatic luminosity $L_{\rm 2~keV}$ to the rest-UV (2500~\AA) monochromatic luminosity $L_{2500}$ defined as $\alpha_{\rm OX} = 0.384\times\log\left(L_{\rm 2~keV} / L_{2500}\right)$.
Figure~\ref{fig;alphaOX} indicates that the red core has a $\alpha_{\rm OX}$ a little smaller than the relation by \cite{Lusso2016}.

Note that the above $L_{2500}$ estimation does not consider the contribution of the host galaxy emission, which may lead to an overestimation of $L_{2500}$.
Therefore, the estimated $L_{2500}$ should be treated as the upper limit.
Figure~\ref{fig;alphaOX} also shows the changes in $\alpha_{\rm OX}$ when $L_{2500}$ is lower than the estimated value, indicating that CID-931 would become closer to the relation by \cite{Lusso2016} when $L_{2500}$ is overestimated.
Accurately estimating the host galaxy contribution with the currently available data is challenging, and it requires spectroscopic data of the red core that can be obtained through future IFU observations.
We also need deeper X-ray observation to constrain $N_{\rm H}$ and decrease the uncertainty of $\alpha_{\rm OX}$.

\subsection{A massive clumpy star-forming host galaxy}\label{ss:host}
Based on the SED fitting of the photometry of the entire complex with a galaxy + AGN template (section~\ref{ss:sed_fitting}), we infer that the stellar mass of CID-931 is $\log \left(M_*/M_\odot\right) = 11.1 \pm 0.1$, and the total SFR is $\left(2\pm1\right)\times10^2M_\odot~{\rm yr^{-1}}$.
These results indicate that CID-931 is on the star-forming main sequence (SFMS) at $z\sim5$ (e.g., \citealt{Salmon2015, Khusanova2021}).
The stellar mass function by \cite{Weaver2023} predicts that only $\sim1$ galaxy with $\log \left(M_*/M_\odot\right) = 11.1 \pm 0.1$ is expected to fall within the COSMOS-Web coverage at $z\sim4.5-5.5$.
This result indicates that CID-931 lies within the most massive galaxy aggregation at $z\sim5$, which is likely found thanks to the wide coverage of COSMOS-Web.

From pixel-by-pixel SED fitting results, we obtain the total stellar mass of $\log\left(M_*/M_\odot\right) = 10.7\pm 0.2$ and total SFR of $\left(2.6\pm0.3\right)\times10^2M_\odot~{\rm yr^{-1}}$.
Since we remove and mask the red core region in the pixel-by-pixel SED fitting (section~\ref{ss:sed_fitting}), the difference in the estimated values should arise from the red core part.
Assuming this, the stellar mass difference of $\log\left(\Delta M_*/M_\odot\right) = 10.9 \pm 0.2$ would correspond to the stellar mass of the red core, which is consistent with the SED fitting results of only the red core ($\log\left(M_*/M_\odot\right) = 10.9 \pm 0.2$, section~\ref{ss:agn}).
There is a possibility that a spatially unresolved SED fitting is biased by the young stellar population (outshining problem, e.g., \citealt{Maraston2010, Sorba2018, Gimenez2023, Gimenez2024, Narayanan2024}), and directly comparing the results obtained from spatially unresolved and resolved methods might be problematic.
Nevertheless, both methods consistently yield $M_* \sim 10^{11} M_\odot$ and ${\rm SFR} \sim$ a few hundred $M_\odot~{\rm yr^{-1}}$, indicating that the CID-931 system is a massive star-forming galaxy (SFG) on the $z\sim5$ SFMS.

\subsubsection{Individual clump properties}\label{ss:clumps}

\begin{figure*}
 \begin{center}
  \includegraphics[width=17cm]{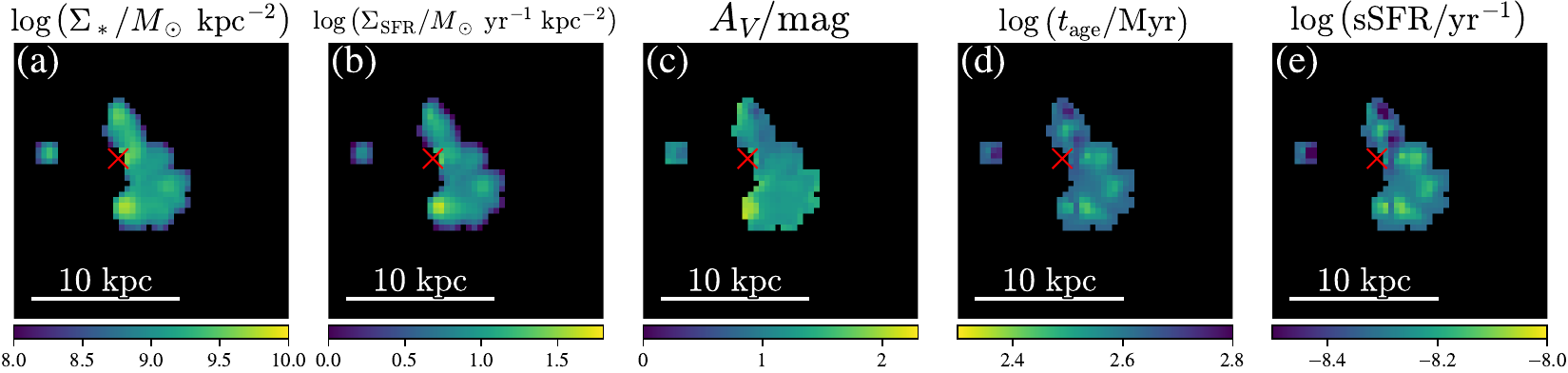} 
 \end{center}
\caption{
Two-dimensional maps of each parameter from pixel-by-pixel SED fitting:
(a) stellar mass density $\Sigma_*$,
(b) SFR density $\Sigma_{\rm SFR}$,
(c) $V$-band dust attenuation $A_V$,
(d) starting time of star-formation activity in the delayed-$\tau$ SFH, $t_{\rm age}$,
and (e) specific SFR.
Note that the PSF component fitted to the red core is subtracted from the input imaging data, and the position of the red core is marked as a red cross.
We see variations between clumps for each parameter.
}
\label{fig;2dsed}
\end{figure*}

Clumpy structures are one of the important morphological features observed in high-$z$ star-forming galaxies (e.g., \citealt{Elmegreen2009, Hainline2023, Kalita2024, Faisst2024, Harikane2024}).
Previous studies have demonstrated that clumpy structures represent rapid star formation triggered by violent disk instabilities (e.g., \citealt{Dekel2009, Dekel2009ApJ, Bournaud2011, Fujimoto2024}) or merger-triggered growth (e.g., \citealt{Jones2024, Nakazato2024, Harikane2024}).

To discuss the characteristics of each clump, we focus on the most luminous eight components.
Figure~\ref{fig;2dsed} shows the distribution of stellar mass density $\Sigma_*$ (panel a), SFR density $\Sigma_{\rm SFR}$ (panel b), $A_V$ (panel c), the age of star-formation activity in the delayed-$\tau$ SFH, $t_{\rm age}$  (panel d), and specific SFR (${\rm sSFR}={\rm SFR}/M_*$, panel e) for the eight clumps reconstructed from the pixel-by-pixel SED fitting results.
Clumps have higher $\Sigma_*$ compared to regions without clumps, reaching up to $10^{10}M_\odot~{\rm kpc^{-2}}$.
$\Sigma_{\rm SFR}$ is also higher in the clumps, exceeding $10^{1.5}M_\odot~{\rm yr^{-1}~{\rm kpc^{-2}}}$.
Additionally, the clumps tend to have smaller $t_{\rm age}$ and larger sSFR.
This tendency indicates clumps contain recently formed stellar components within the CID-931 system.
The $A_V$ map reveals that the clump in the southern part of CID-931, which has the highest $\Sigma_*$ and $\Sigma_{\rm SFR}$, has a significant dust attenuation with $A_V>2$, also consistent with the red color observed in the three-color image in figure~\ref{fig;imgs}.

\begin{figure*}
 \begin{center}
  \includegraphics[width=15cm]{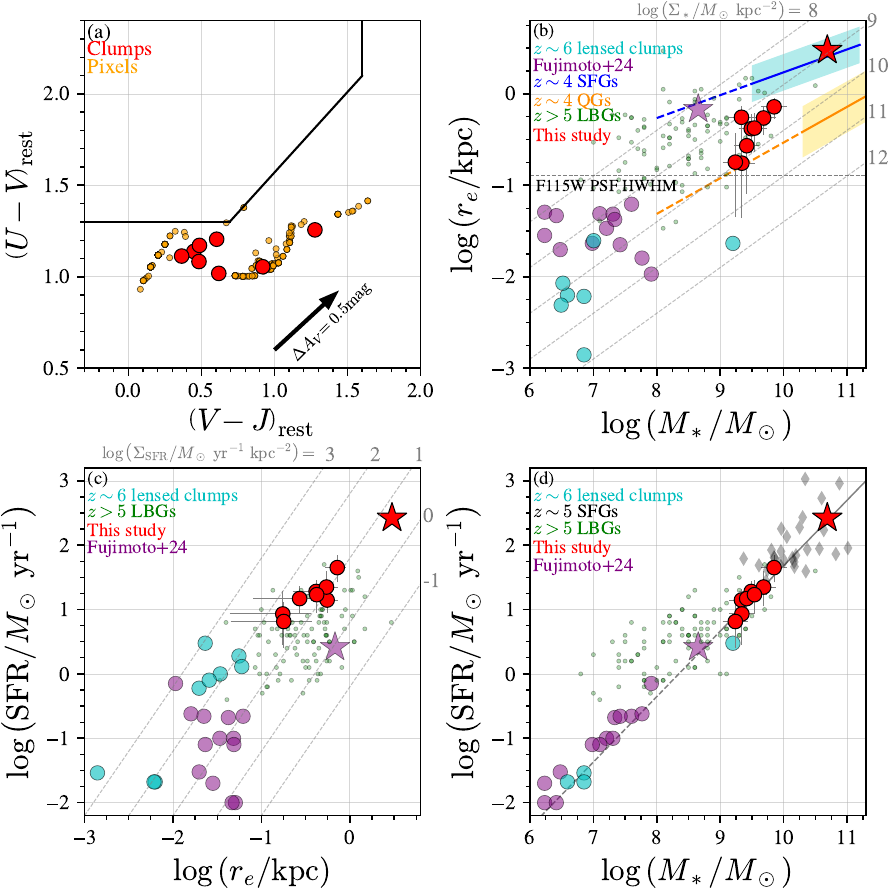} 
 \end{center}
\caption{
Galaxy and clump characteristics.
(a) Rest-frame $UVJ$ color-color diagram.
Orange and red dots indicate each pixel and clump, respectively.
The black line indicates the threshold for classifying QG and SFG by \cite{Williams2009}.
The black arrow indicates the dust extinction corresponding to $\Delta A_V=0.5~{\rm mag}$.
(b) Size ($r_e$) - mass ($M_*$) relation.
Red and purple dots represent each clump in CID-931 and $z\sim6$ clumpy system dubbed ``Cosmic Grapes'' in \cite{Fujimoto2024}, respectively.
Stars indicate the overall region of CID-931 and Cosmic Grapes (note that the red core component is not included in CID-931).
Cyan dots indicate $z\gtrsim6$ lensed clumps (\citealt{Welch2023, Vanzella2023, Fudamoto2024}).
Green dots represent high-$z$ Lyman break galaxies (LBGs, \citealt{Morishita2024}).
The blue and orange lines and shaded regions indicate the size-mass relation for $z\sim4$ SFGs \citep{Ward2024} and UVJ-selected QGs \citep{Ito2024}.
The dashed lines indicate the extrapolation of each size-mass relation.
The horizontal gray dashed line represents the half width at half maximum (HWHM) of the F115W PSF, while the diagonal gray dashed lines indicate fixed values of  $\Sigma_*$.
(c) Size ($r_e$) - SFR relation.
Each marker is the same as panel~(b).
The diagonal gray dotted lines indicate each $\Sigma_{\rm SFR}$.
(d) SFR - $M_*$ relation.
Each marker is the same as panel~(b).
The gray diamonds indicate $z\sim5$ SFGs \citep{Khusanova2021, Xiao2023}.
The black solid and dashed lines represent the $z\sim5$ SFMS \citep{Khusanova2021} and its extrapolation.
The total $M_*$ and SFR (red stars) are the sum of the values of each pixel.
For the total $r_e$, we use $r_e$ for the HSC $i$-band (section~\ref{ss:sample}).
Each clump in CID-931 is on the $z\sim5$ SFMS, massive with respect to the JWST SFG population, and more compact with higher $\Sigma_*$ and $\Sigma_{\rm SFR}$ than typical high-$z$ SFGs.
}
\label{fig;galaxy}
\end{figure*}

Figure~\ref{fig;galaxy} presents the UVJ diagram\footnote{We calculated the model magnitude based on the best-fit SED with CIGALE SED fitting code. Due to the discrete model grid, the color indices are discrete, and some pixels have the same color indices in figure~\ref{fig;galaxy}a.} and the relationships between size, SFR, and $M_*$ for each clump reconstructed from the 2D decomposition results.
Firstly, each pixel and clump is classified as an SFG in the UVJ diagram (e.g., \citealt{Williams2009}).
Each clump has a high stellar mass ($10^9<M_*<10^{10}M_\odot$), which is among the most massive clumps ever found (e.g., \citealt{Elmegreen2009, Huertas2020, Ambachew2022}).
The SFR-$M_*$ relation shows that each clump lies on the $z\sim5$ SFMS, indicating they are star-forming clumps.

Compared to the $z\sim4$ size-mass relation of SFGs \citep{Ward2024} based on JWST measurements, these clumps are approximately ten times smaller in $r_e$ than $z\sim4$ SFGs with similar $M_*$.
The clumps exhibit $\Sigma_*\sim10^{9-10}M_\odot~{\rm kpc^{-2}}$ consistent with or little higher than the extrapolation of the $z\sim4$ size-mass relation of UVJ-selected quiescent galaxies (QGs, \citealt{Ito2024}).
Since galaxy size is wavelength-dependent and the size-mass relations of \cite{Ward2024} and \cite{Ito2024} are based on $r_e$ at rest-frame 5000~\AA, we use $r_e$ measurements in the F277W filter which is the closest to the rest-frame 5000~\AA\, at $z=4.91$.
However, the total $r_e$, indicated by the star in figures~\ref{fig;galaxy}b and c, is based on HSC $i$-band measurements introduced in section~\ref{ss:sample}.
We do not correct for wavelength dependence for the total $r_e$ because CID-931 is not a single-galaxy system but a complex one that may not follow the typical size-wavelength relation.
Even if we applied a correction using the size-wavelength relation from \cite{Ito2024}, the total $r_e$ would change only from $r_e \sim 0\farcs5$ to $\sim 0\farcs3$ without significantly altering its position on the $r_e$-$M_*$ or -SFR planes.

When compared to the spectroscopically confirmed $z>5$ galaxies \citep{Morishita2024}, $M_*$ and SFR of our clumps are comparable with the most massive or highest-SFR galaxies and also show high values of $\Sigma_*\sim10^{9-10}M_\odot~{\rm kpc^{-2}}$ and $\Sigma_{\rm SFR}\sim10^{1-2}M_\odot~{\rm yr^{-1}~kpc^{-2}}$.
These $\Sigma_*$ and $\Sigma_{\rm SFR}$ are as high as those of gravitationally lensed clumps at $z\sim6$ \citep{Fujimoto2024, Gimenez2024, Valentino2024}.
However, comparing the clumps in CID-931 and those reported in \cite{Fujimoto2024}, there is a significant scale difference of approximately ten times in $r_e$, suggesting different types of objects.

Furthermore, \cite{Dessauges2017} reported that observational limitations on spatial resolution lead to a lower limit of the detectable clump $M_*$.
Also, the clumpy structures seen in \cite{Fujimoto2024} could not be resolved without gravitational lensing (see figure~2 in \citealt{Fujimoto2024}).
Similarly, if the clumps in CID-931 are composed of clumps with smaller stellar masses with a separation smaller than spatial resolution, the absence of gravitational lens magnification would prevent us from resolving smaller clumps.
If the observed clumps are the aggregation of smaller clumps, these smaller clumps should have a higher $\Sigma_{\rm SFR}$ and $\Sigma_*$ than the observed clumps.
Assuming each clump within the CID-931 system is further divided into smaller clumps with similar $M_*$ distribution ratios and $\Sigma_*\sim10^{11}M_\odot~{\rm kpc^{-2}}$, the smaller clumps would correspond to massive clusters, such as the Galactic young massive clusters or nuclear star clusters \citep{Norris2014}, with masses of approximately $10^7-10^9M_\odot$ and sizes of $r_e\sim4-40~{\rm pc}$.
Such compact, dense clumps would be similar to the $z\gtrsim6$ clumps previously observed thanks to gravitational lensing (e.g., \citealt{Welch2023, Vanzella2023, Fudamoto2024, Fujimoto2024}) and can be analogs of young massive star clusters (YMC, e.g., \citealt{Portegies2010}) in the local Universe and usually with pc-scale size.

\subsection{Comparison with zoom-in simulations}\label{ss:simu}
Despite CID-931 having interesting features of a massive clumpy structure, the currently available observational data for CID-931 are limited. 
Especially, the lack of high spatial resolution spectroscopic data from ALMA or JWST IFU prevents us from discussing the kinematics of each clump within the galaxy. 
To discuss the possible formation mechanism of the observed clumpy structures, we use zoom-in cosmological simulations FirstLight \citep{Ceverino2017}. 
The simulation suite has a box size of 60 comoving Mpc, and the maximum spatial resolution is 17 pc, which is well suited to studying inner structures of massive rare galaxies such as CID-931.
Note that our simulation does not include AGN feedback.
In this section, we discuss the formation mechanism of massive clumpy galaxies at high redshift, assuming that the AGN feedback is negligible for the formation of clumpy morphologies. 

Using the clumpy identification of \cite{Nakazato2024}, we select clumpy galaxies with $M_* > 10^{10} M_\odot$ at $z < 6$ snapshot and successfully find an analog of the clumpy galaxies with the stellar mass of $M_*= 10^{10.1}M_\odot$ at $z=5.94$.
Although the simulated analog is about ten times less massive than CID-931, it is still considered massive galaxy at $z\sim6$, similar to CID-931.
Thus, we regard it as an analog of a massive clumpy galaxy at high redshift.

\begin{figure*}
 \begin{center}
  \includegraphics[width=17cm]{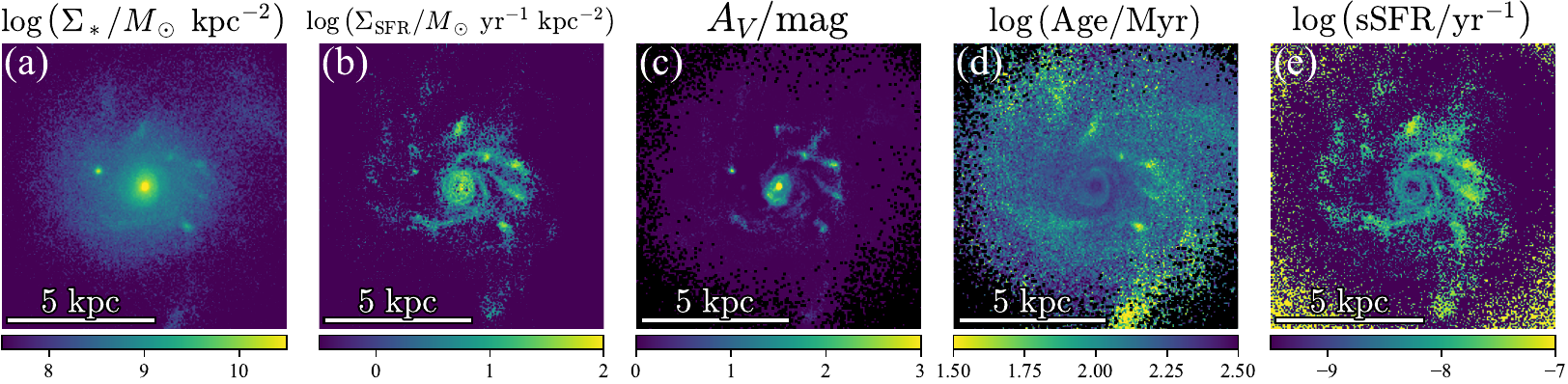} 
 \end{center}
 \caption{
 Same for figure~\ref{fig;2dsed}, but for the simulated galaxy.
 }
\label{fig;simu_param}
\end{figure*}

Panels (a) and (b) of figure~\ref{fig;simu_param} show the surface density map of stellar mass ($\Sigma_*$) and SFR ($\Sigma_{\rm SFR}$).
We see that clumps have $\Sigma_*\sim10^{9-10}M_\odot~{\rm kpc^{-2}}$ and $\Sigma_{\rm SFR}\sim10^{1-2}M_\odot~{\rm yr^{-1}~kpc^{-2}}$, which are consistent with those of CID-931 (figure~\ref{fig;2dsed}).
Dust attenuation of the simulated galaxies is calculated by using post-processing radiative transfer code {\tt SKIRT} \citep{Beas2011, Camps2020} in the same manner as in \cite{Behrens2018} and Nakazato et al. (in preparation).
Panel (c) of figure~\ref{fig;simu_param} showing the $V$-band attenuation map indicates that the clumps have large-$A_V$ of $A_V\gtrsim2$, as similar to CID-931.
Panels (d) and (e) of figure~\ref{fig;simu_param} exhibit the distribution of mass-weighted age and specific SFR (${\rm sSFR} \equiv \Sigma_{\rm SFR}/\Sigma_*$).
The central core has an age of $\sim$120 Myr and sSFR of $\lesssim 1~{\rm Gyr^{-1}}$, while the surrounding clumps are much young ($\lesssim$50~Myr) and proceed bursty star formation with sSFR $\gtrsim10~{\rm Gyr^{-1}}$.

\begin{figure}
 \begin{center}
  \includegraphics[width=8cm]{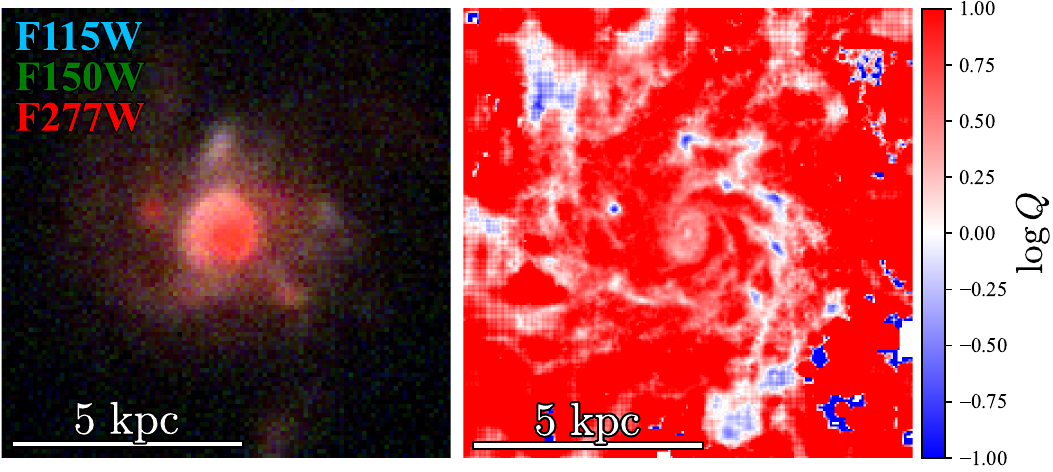} 
 \end{center}
 \caption{
 (Left) Mock three-color (JWST/NIRCam F277W, F150W, and F115W for RGB) image of the simulated galaxy with $\log\left(M_*/M_\odot\right)\sim10.1$ at $z\sim5.94$.
 We convolve the image assuming a Gaussian PSF that corresponds to the physical size of the F115W PSF at $z\sim4.91$.
 (Right) Toomre-$Q$ map.
 Blue and red indicate $Q<1$ and $Q>1$, respectively.
 The simulated galaxy similar to CID-931 suggests that the clumpy structure of these galaxies is formed through disk instability. 
 }
\label{fig;simu}
\end{figure}

We further generate the mock JWST three-color images as shown in the left panel of figure~\ref{fig;simu}.
The three filters of F115W, F150W, and F277W are combined and the corresponding PSF and transmission are considered.
The mock image shows a red core and separated clump regions similar to CID-931.
The 3-color mock image also suggests that AGNs in complex clumpy systems may appear as offset AGNs \citep{Li2024} with separated other components when their clumps are significantly brighter than the disk and have asymmetric distributions.

Regarding the clump formation processes, we check the snapshots preceding the one shown in figure~\ref{fig;simu} and confirm that there are no merger events to trigger the clump formation.
We also calculate Toomre-$Q$ parameter \citep{Toomre1964}, defined as \begin{equation}
    Q = \frac{\sigma \kappa}{AG\Sigma}, \label{eq:Toomre}
\end{equation}
where $\sigma$, $\kappa$, and $\Sigma$ are the radial velocity dispersion, the epicyclic frequency of a disk, and the surface density, respectively.
$G$ and $A$ are the gravitational constant and numerical factor ($A=\pi$ for a gas disk and $A\simeq3.36$ for a stellar disk, see \citealt{Toomre1964, Elmegreen2011, Inoue2016}).
When $Q\gg1$, the disk is in a stable state, and the effects of self-gravity are neglectable. 
Otherwise, $Q<1$ means that gravitational instability occurs with axisymmetric perturbation, leading to clump formation and disk fragmentation. 
Note that \cite{Inoue2016} suggest that some protoclumps may show $Q > 1$ and low Toomre-$Q$ does not fully trace clump formation through disk instability.

We calculate the Toomre-$Q$ map assuming a two-component disk of gas and stars in the same manner as \cite{Inoue2016} and \cite{Ceverino2010}, and present the face-on view in the right panel of figure~\ref{fig;simu}.
We see the apparent disk structure with $Q > 1$ and clumps with $Q < 1$, suggesting clump formation through violent disk instability.
Therefore, the zoom-in simulation results suggest the possibility that violent disk instability can form massive clumpy structures in the high-$z$ massive galaxies.

\section{Discussion}\label{s:discussion}

\subsection{Formation mechanism of the clumpy structure}\label{ss:cause_clumpy}
The physical origins of these clumpy structures are essential to understanding the rapid galaxy formation and the relation with the AGN activity in the early Universe.
However, it is still unclear which processes significantly contribute to the clump formation, mergers or disk instabilities.

Explaining all clumps in CID-931 as merger-induced structures needs a highly complex merger event of multiple massive SFGs, at least eight galaxies with $M_*\sim10^9-10^{10}M_\odot$.
\cite{Jin2023} report such a complex merger event of massive galaxies, a $z\sim5$ compact group of galaxies.
The stellar mass range of each member galaxy is $M_*\sim10^{8.4-9.8}M_\odot$, and \cite{Jin2023} discuss that the member galaxies will merge into a single massive ($M_*\sim10^{11}$) galaxy at $z\sim3$ based on the simulations.
Compared with this compact group of galaxies, CID-931 is a little more compact than the object in \cite{Jin2023} (spread over $10\times20~{\rm kpc^{2}}$) and may represent a stage where the member galaxies start accreting into the center and are about to merge.
During such a period, the gas inflow into the central region caused by the interaction may trigger the AGN activity.
This scenario is consistent with the galaxy-SMBH evolution paradigm by \cite{Hopkins2008}.

The disk instability scenario suggests that star-forming clumps are formed through gravitational disk instability, possibly driven by gas inflow (e.g., \citealt{Bournaud2011}).
It is worth noting that the $z\sim6$ clumpy system in \cite{Fujimoto2024}, which follows similar scaling relations to CID-931 (figure~\ref{fig;galaxy}), is also suggested to be formed through disk instability from Toomre-$Q$ estimation based on ALMA observations.
As described in section~\ref{ss:simu}, zoom-in simulations find a galaxy with similar morphologies and clump properties to CID-931 and show that the clumps are formed by disk instabilities.
In the disk instability scenario, the gas inflow may also supply dust and gas into the central region and initiate dusty AGN activity.

To identify the origin of the clumpy structure, we need to measure the kinematics of each clump and gas using spatially resolved spectroscopic data and evaluate disk instability by calculating Toomre-$Q$.
Therefore, future high-spatial resolution and deep integral field spectroscopy with ALMA and JWST are necessary.

As discussed in section~\ref{ss:clumps}, each clump might have multiple unresolved clumps that even JWST can not spatially resolve.
These components would be resolved using higher spatial resolution observations using ALMA or next-generation large telescopes, such as next generation Very Large Array (ngVLA), Extremely Large Telescope (ELT), and Thirty Meter Telescope (TMT).

\subsection{SMBH-galaxy relation}\label{ss:mass_relation}

Due to the lack of a near-infrared spectrum, we cannot measure the $M_{\rm BH}$ of the red core from the present photometric data. 
Here, we estimate $M_{\rm BH}$ by assuming the Eddington limit, i.e., the maximum accretion speed.
We convert the X-ray luminosity of CID-931 (section~\ref{ss:xray}) to bolometric luminosity $L_{\rm bol} = \left(6.3\pm1.5\right)\times10^{45}~{\rm erg~s^{-1}}$ by assuming a bolometric correction presented in \cite{Duras2020}. 
With the assumption of the Eddington limit accretion, the corresponding black hole mass is $M_{\rm BH} = \left(4.4\pm1.2\right)\times10^{7}M_\odot$ that should be regarded as a lower limit.
Note that the bolometric correction factor can be different for high-$z$ AGNs, as suggested in recent JWST studies (e.g., \citealt{Maiolino2024_chandra, Juodzbalis2024}).

\begin{figure}
 \begin{center}
  \includegraphics[width=8cm]{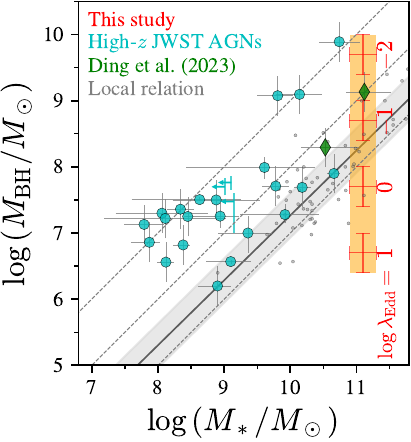} 
 \end{center}
\caption{
The relation between $M_{\rm BH}$ and $M_*$.
Cyan, green, and gray plots indicate the high-$z$ JWST discovered AGNs \citep{Harikane2023, Maiolino2023, Yue2023}, $z\sim6$ low luminosity quasars \citep{Ding2022_z6}, and the local active \citep{Bennert2011} and inactive galaxies \citep{Haring2004}. 
The orange rectangle shows the possible positions of CID-931 with different Eddington ratio $\lambda_{\rm Edd}$.
Gray dashed and black solid lines indicate the mass ratio $\log\left(M_{\rm BH}/M_*\right)=-3, -2, -1$ and the local mass relation obtained from the fitting of the samples in \cite{Haring2004} and \cite{Bennert2011}.
}
\label{fig;mm}
\end{figure}

The estimated $M_*$ and $M_{\rm BH}$ indicate a system with higher $M_*$ than those found by JWST among LRDs and high-$z$ AGNs (figure~\ref{fig;mm}).
Meanwhile, J2236+0032, a low-luminosity quasar at $z=6.40$ reported in \cite{Ding2022_z6}, has a similar masses of $\log\left(M_*/M_\odot\right)={11.12^{+0.40}_{-0.27}}$ and $M_{\rm BH}=\left(15.4\pm2.7\right)\times10^8M_\odot$.
Also note that some high-$z$ quasars are estimated at similar dynamical masses $M_{\rm dyn}\sim 10^{11}M_\odot$ (e.g., \citealt{Izumi2019, Neelman2021, Izumi2021}). 

In the widely accepted BH-galaxy evolution paradigm (e.g., \citealt{Hopkins2008}), dusty AGNs are in a transitional phase evolving into unobscured quasars by expelling dust.
Therefore, CID-931 potentially evolves into high-$z$ quasars reported in the previous studies with $M_*$ or $M_{\rm dyn}\sim10^{11}M_\odot$.
Additionally, simulation studies have suggested that high-$z$ giant and high-$\Sigma_*$ clumps accumulate at the center to form a bulge (e.g., \citealt{Elmegreen2008, Ceverino2010, Ceverino2012, Bournaud2014, Mandelker2014, Mandelker2017}). 
CID-931 may also evolve into a massive elliptical galaxy as its clumps coalesce to form a bulge.
We can verify this scenario through spectroscopic follow-up observations by exploring outflows to discuss AGN feedback and constraining SFR and depletion time to discuss future mass growth.

The clumpy structure of CID-931 suggests that its AGN activity is associated with the active star formation or the complex merger event (section~\ref{ss:cause_clumpy}).
\cite{DeGraf2017} also suggested that such clumps accrete into the central region and contribute to the strong outflow by fueling the AGN.
Following this scenario, future accretion of the currently observed massive clumps in CID-931 may trigger the AGN feedback to expel the dust and the evolution into an unobscured quasar.
Systematic studies of dusty AGN host galaxies are required to deepen our understanding of the relationship between AGN activity and host galaxy evolution in the transitioning phase.

\subsection{Comparisons with JWST's LRDs}\label{ss:LRD_compare}
LRD studies (e.g., \citealt{Greene2024}) have reported very red colors and a V-shaped SED characterized by a change in the spectral slope around $\lambda_{\rm rest}\sim 3000-4000$~\AA.
In the case of CID-931, the slope observed in F115W and F150W ($\lambda_{\rm rest}\sim 2000-2500$~\AA) is similar to the slope in F277W and F444W (rest-frame $\lambda_{\rm rest}\sim 4700-7500$~\AA).
Due to the lack of the presence of the V-shape SED for the red core, we do not conclude that CID-931 is the same type of object as existing LRDs.
However, given the uniqueness of CID-931 as a high-$z$ X-ray-detected dusty AGN with similar spectral slope in rest-optical to LRDs, it is still worth comparing CID-931 with existing LRDs in the literature to better understand the nature of both CID-931 and LRDs.

Most of the previously-found LRDs are undetected in X-ray (\citealt{Kocevski2023, Furtak2023, Matthee2024}) even after a stacking analysis \citep{Yue2024, Ananna2024, Maiolino2024_chandra, Akins2024} except for only two cases among hundreds of LRDs reported as X-ray detected \citep{Kocevski2024}. 
Interestingly, the two X-ray-detected LRDs in \cite{Kocevski2024} have a massive ($M_*\sim10^{10.6}M_\odot$) stellar component (PRIMER-COS 3866 with $L_{\rm bol}\sim1\times10^{46}~{\rm erg~s^{-1}}$ at $z_{\rm spec}=4.66$) or an extended blue clumpy structure (JADES 21925 with $L_{\rm bol}\sim1\times10^{45}~{\rm erg~s^{-1}}$ at $z_{\rm photo}=3.1$)\footnote{The bolometric luminosity are calculated from the bolometric correction by \cite{Duras2020}}.
In section~\ref{ss:xray}, we also find that CID-931 has a $N_{\rm H}$ upper limit of $N_{\rm H}\lesssim8\times10^{22}~{\rm cm^{-2}}$ and is not a Compton-thick AGN ($N_{\rm H}\gtrsim10^{24}~{\rm cm^{-2}}$).
This $N_{\rm H}$ upper limit is also consistent with both of the two X-ray-detected LRDs from \cite{Kocevski2024} with $N_{\rm H}\sim10^{23}~{\rm cm^{-2}}$.
Therefore, compared to typical X-ray-undetected LRDs that usually lack host galaxy components, high-$z$ X-ray-detected LRDs or dusty AGNs (\citealt{Kocevski2024} and this study) might represent a different population with extended host galaxy components.

For further and solid discussion, it is crucial to enlarge the sample of similar X-ray- or host-detected high-$z$ dusty AGNs to perform a statistical abundance comparison and make spectroscopic observations to compare with typical LRDs.
Future deeper X-ray telescopes such as Athena can enlarge the sample size of high-$z$ X-ray-detected AGNs.

\section{Conclusions}\label{s:conclusion}

Applying the pixel-by-pixel color selection method on the JWST high spatial-resolution imaging data obtained with the COSMOS-Web program, we discover a red unresolved component surrounded by a clumpy structure in CID-931, an X-ray-detected AGN at $z_{\rm spec}=4.91$.
CID-931 is a unique high-$z$ system with X-ray detection as solid evidence of an AGN and a very complex clumpy morphology, which we can directly detect without complicated imaging decomposition analysis. 
We conduct a detailed analysis of CID-931 based on the JWST imaging data.

Image-based decomposition indicates that the red core, the brightest component in the F277W and F444W images, shows a rest-optical color similar to those of high-$z$ dusty AGNs and LRDs (figure~\ref{fig;core_SED}) and suggests it is a dusty AGN. 
Analysis of the X-ray flux shows that the AGN in CID-931 is not a Compton-thick AGN ($N_{\rm H}\lesssim8\times10^{22}~{\rm cm^{-2}}$ corresponding to the limiting depth in the Chandra observation), similar to the X-ray-detected LRDs reported by \cite{Kocevski2024}.
We also find the red core has a large bolometric luminosity of $L_{\rm bol}\sim 6.3\times10^{45}~{\rm erg~s^{-1}}$.

SED fitting revealed that the entire CID-931 is a massive system with $M_*\sim10^{11}M_\odot$, similar to the mass observed in high-$z$ quasars (e.g., \citealt{Izumi2019, Ding2022_z6}).
We roughly estimate the $M_{\rm BH}$ from the bolometric luminosity and find that CID-931 has a significantly larger stellar mass than existing JWST-found AGNs and occupies a different parameter space on the $M_{\rm BH}-M_*$ plane (figure~\ref{fig;mm}).

CID-931 has at least eight massive star-forming clumps over $1.\!^{\prime\prime}6 \approx 10~{\rm kpc}$.
Although each clump has a high $M_*$ ($\sim10^{9-10} M_\odot$) and SFR ($\sim10^{1-2} M_\odot~{\rm yr^{-1}}$) comparable with high-$z$ Lyman break galaxies, they have much more compact morphology, suggesting higher $\Sigma_*$ ($\sim10^{9-10}M_\odot~{\rm kpc^{-2}}$) and $\Sigma_{\rm SFR}$ ($\sim10^{1-2}M_\odot~{\rm yr^{-1}~kpc^{-2}}$) as shown in figure~\ref{fig;galaxy}.

We discussed the two possibilities of the formation mechanism of the CID-931 clumpy structure.
One scenario is that the massive clumps are massive galaxies that are accreting and coalescing into a single massive galaxy.
The other scenario is that gas inflows cause the clump formation through a violent disk instability, as suggested by zoom-in simulations (figure~\ref{fig;simu}).
Such a merger event or gas inflow might have also triggered the AGN activity by supplying gas into the central region.

We propose the possibility that the X-ray-detected LRDs, all of which have host galaxy components, represent a different population from previously discovered LRDs.
Also, considering that typical LRDs usually lack an X-ray detection and a host galaxy component, X-ray-detected and host-detected LRDs might bridge between typical LRDs and existing SMBH + galaxy systems. 
To discuss the role of such objects in the galaxy-SMBH evolution more solidly, it is essential to increase the sample of X-ray-detected or host galaxy-detected high-$z$ AGNs, discuss their statistical properties, and perform abundance comparisons.
For this purpose, utilizing spatial information is useful because it can select host-detected dusty AGNs that are overlooked in typical selection methods (see Tanaka et al. in preparation).

For a more detailed evaluation of SMBH activity, including spectroscopic identification of the AGN, measurement of $M_{\rm BH}$, and exploration of AGN outflows to study AGN feedback, spatially-resolved and deep rest-optical follow-up observation with JWST/NIRSpec IFU is needed.
Future high spatial and velocity resolution observations by ALMA could measure the kinematics of each clump within the host galaxy, enabling the calculation of the Toomre-$Q$ parameter to assess whether the host galaxy accumulates mass through mergers or disk instability. 
Additionally, future ultra-high spatial resolution observations using ngVLA, ELT, and TMT are needed to spatially resolve finer clump structures that even JWST cannot.
Also, deeper X-ray observations are needed to strongly constrain $N_{\rm H}$ and $\Gamma$, possibly with future deeper X-ray telescope surveys, such as Athena.

\begin{ack}
We thank Kei Ito, Makoto Ando, and Suin Matsui for the fruitful discussion.
We thank the anonymous referee for helpful feedback.
This work is based on observations made with the NASA/ESA/CSA James Webb Space Telescope.
The data were obtained from the Mikulski Archive for Space Telescopes at the Space Telescope Science Institute, which is operated by the Association of Universities for Research in Astronomy, Inc., under NASA contract NAS 5-03127 for JWST.
These observations are associated with program IDs 1727 and 1837.
Numerical computations were in part carried out on the iDark cluster, Kavli IPMU and the analysis servers at Center for Computational Astrophysics, National Astronomical  Observatory of Japan.
This work was made possible by utilizing the CANDIDE cluster at the Institut d’Astrophysique de Paris, which was funded through grants from the PNCG, CNES, DIM-ACAV, and the Cosmic Dawn Center and maintained by Stephane Rouberol.
\end{ack}

\section*{Funding}
Kavli IPMU is supported by World Premier International Research Center Initiative (WPI), MEXT, Japan.
TT is supported by the Forefront Physics and Mathematics Program to Drive Transformation (FoPM), a World-leading Innovative Graduate Study (WINGS) Program at the University of Tokyo.
YN acknowledges funding from JSPS KAKENHI Grant Number 23KJ0728 and a JSR fellowship.
YF acknowledge support by JSPS KAKENHI Grant Numbers JP22K21349 and JP23K13149.
SJ is supported by the European Union's Horizon Europe research and innovation program under the Marie Sk\l{}odowska-Curie grant No. 101060888.

\section*{Data availability} 

\bibliographystyle{plainnat2}
\bibliography{export}{}

\end{document}